\documentclass{aa}  

\usepackage{graphicx}
\usepackage{txfonts}
\usepackage{xspace}
\usepackage{physics}
\usepackage{changes}
\definechangesauthor[name=Mitya, color=orange]{DK}

\usepackage{hyperref}

\newcommand{\ls}{LS~5039\xspace}
\newcommand{\lsi}{LS~I~+61 303\xspace}
\newcommand{\psr}{PSR~B1259-63/LS2883\xspace}
\newcommand{\hess}{H.E.S.S.\xspace}

\begin{document}

   \title{The production of orbitally modulated UHE photons in \ls}

\titlerunning{UHE emission in the $\gamma$-ray binary LS~5039}

   \author{V. Bosch-Ramon\inst{1}
        \and D. Khangulyan\inst{2,3}
        }

        \institute{Departament de F\'{i}sica Qu\`antica i Astrof\'{i}sica, Institut de Ci\`encies del Cosmos (ICC), Universitat de Barcelona (IEEC-UB), Mart\'{i} i Franqu\`es 1, E08028 Barcelona, Spain\\
        \email{vbosch@fqa.ub.edu}
        \and
          Key Laboratory of Particle Astrophysics, Institute of High Energy Physics, Chinese Academy of
     Sciences, 100049 Beijing, People's Republic of China
     \and
      Tianfu Cosmic Ray Research Center, 610000 Chengdu, Sichuan, People's Republic of China
            }

  \abstract
   {Gamma-ray binaries present emission that is variable and can reach ultra-high energies. The processes behind the acceleration of the particles that produce this very energetic radiation are yet to be understood.}
   {We probe the properties of the particle accelerator and the ultra-high-energy photon emitter in the gamma-ray binary \ls.} 
   {From the properties of the binary system and the ultra-high-energy radiation detected by HAWC, we used analytical tools to investigate how these properties constrain the emission and acceleration regions, namely the role of synchrotron losses, particle confinement, and the accelerated particle spectrum, and propose an acceleration scenario that can relax the derived constraints.}
   {The modest target densities for hadronic processes and the overall gamma-ray orbital variability favor inverse Compton scattering of ultraviolet photons from the massive companion star by highly relativistic electrons. The acceleration of the highest energy electrons implies a constraint on synchrotron cooling in the acceleration region, which can set an upper limit on its magnetic field. Moreover, the detected variability requires very strong particle confinement in both the acceleration and emission regions, which sets a lower limit on their magnetic fields that is barely consistent with the synchrotron cooling constraint from acceleration. Synchrotron losses may be higher in the emitting region if it is separated from the accelerator, but this requires a very hard particle injection spectrum. An accelerator based on an ultrarelativistic magnetized outflow can alleviate these requirements.}
   {A scenario for \ls of the kind proposed by Derishev and collaborators, in which an ultrarelativistic magnetized outflow accelerates leptons injected within the outflow by \(\gamma\gamma\) absorption, provides a viable mechanism to accelerate very energetic electrons. This mechanism relaxes the acceleration and confinement requirements by reducing the impact of synchrotron cooling, and can generate the required particle spectrum.}


   \maketitle

\section{Introduction}\label{intro}

Twenty years ago, the High Energy Stereoscopic System (\hess) detected orbitally modulated, very-high-energy
(VHE; $100\,{\rm GeV}\lesssim \epsilon\lesssim 100$~TeV) emission and photons with energies reaching $\epsilon\sim 30$~TeV from \ls
 \citep{aha06}, a high-mass binary considered to be a likely GeV-emitting microquasar at the time \citep{par00}. Some years later, Fermi confirmed
the source to be a GeV emitter \citep{fer09,had12}, and a re-analysis of MeV data
strengthened previous evidence of detection in the MeV range \citep{col14}. This
information was complemented by soft and hard X-ray observations, which found an
orbital light curve similar to that in TeV \citep[e.g.,][]{bos05,hof09,tak09}, and
to that hinted by the MeV data re-analysis. The High-Altitude Water Cherenkov Observatory (HAWC) has recently detected the
source up to $\epsilon\approx 200$~TeV, reaching the photon ultra-high-energy (UHE; $\epsilon>100$~TeV) range, with a spectral energy distribution (SED) $\propto \epsilon^{-0.8}$ that shows no evidence of a high-energy cut-off \citep{haw24}. HAWC has also found $2.7\sigma$ evidence of orbital variability in the $40-118$~TeV range, and $2.2\sigma$ evidence above 100~TeV. Furthermore, the emission detected around the inferior conjunction of the compact object (INFC; orbital phase range $0.45-0.9$) reaches 208~TeV, whereas around the superior conjunction (SUPC) the emission reaches 118~TeV, which also suggests variability above 100~TeV.

\ls was initially considered to be a microquasar due to the presence of X-ray
accretion features and radio jets \citep{rib99,par00}. However, several authors
proposed that this system may host a young pulsar instead \citep{mar05,dub06}, as in
the case of \psr, which is also a TeV-emitting binary \citep{joh92,aha05}. In
addition, it was found that the previously found X-ray accretion features were
likely to be background contamination due to the poor angular resolution of {\it
RXTE} \citep{rib99,bos05}. However, while a pulsar-star wind collision scenario is
generally accepted to be behind the nonthermal (NT) emission in \psr (e.g.,
\citealt{tav97,kha07}; see however \citealt{yi17}), in \ls a microquasar scenario
cannot yet be fully discarded. The compact object mass has not been ascertained
\citep{cas05}, the presence of pulsations is controversial
\citep{yon20,kar23,mak23}, and some accretion modes may not yield prominent features
such as thermal emission and spectroscopic lines \citep{oka08,bar12}. More exotic
scenarios have been discussed as well, such as a young magnetar, prompted by the
hint of pulsations with a period of $\approx 9$~s found in X-rays \citep{yon20}.
Such an object, however, seems to be at odds with the absence of a conspicuous
supernova remnant surrounding the source
\citep[][]{rib02,mol12}, and assessing its feasibility needs further theoretical
development (Bosch-Ramon \& Barkov, in prep.). Therefore, the debate about the processes yielding the NT emission in
\ls remains far from being solved, a debate that is tightly connected to other
similar systems. In particular, in \lsi, which presents evidence for 269~ms radio
pulsations \citep{wen22} and thus disfavors the microquasar scenario \citep[see
however][]{jar24}, the question of the origin of the NT emission is still open
\citep[e.g., a colliding-wind model vs. a transitional --~magnetar-like~-- pulsar
scenario; e.g.,][]{dub06,pap12}.

Despite the puzzles yet to be solved, combining data and theory has helped to reach some conclusions on the properties of the accelerator and the NT emitter in \ls. For instance, the multiwavelength NT behavior of \ls, and in particular its gamma-ray orbital variability, strongly suggest inverse Compton (IC) scattering of stellar photons as the origin of this emission \citep[e.g.,][]{kha08,dub08,tak09}. It is worth noting that HAWC results match the behavior found by \hess at an overlapping energy range well, so a different mechanism for the UHE emission seems unlikely. Moreover, the modest mass-loss rate of the stellar companion \citep{cas05} further strengthens that hypothesis, as luminous proton-proton emission or relativistic Bremsstrahlung would require an unrealistic energy budget for either protons or electrons, respectively. In addition, the hard photon local fields are too diluted for efficient photo-meson production \citep{bos09}. \cite{kha08} also showed that the detection of 30~TeV photons by \hess favored an efficient accelerator located in the periphery of the binary to soften the strong acceleration rate requirements, but not far from it due to the presence of variability. Similar conclusions were also reached in \cite{bos07} and \cite{bos08b} for the X-ray and the VHE emitter, respectively, using arguments based on X-ray absorption and gamma-ray reprocessing. These articles also suggested that this class of scenario was more compatible with a jet-like structure, although an alternative conclusion about the X-ray emitter structure was derived by \cite{szo11}, who proposed an extended emitting region with a size similar to the system semimajor axis ($a$), interpreting it as the star-pulsar wind colliding region. Moreover, theoretical studies of the star-pulsar wind interaction along the orbit also predicted the acceleration of particles outside the binary due to a combination of the Coriolis force and the slow stellar wind, which should trigger a strong lateral shock on the pulsar wind as it propagates away from the orbiting binary (the Coriolis shock; e.g., \citealt{bos11,bos12}). Numerical studies also showed that the magnetic field could introduce further structure to the two-wind interaction region on the scales of the binary system \citep[e.g.,][]{bog19,bar24}.

In this work, we aimed to improve our knowledge of the
NT accelerator and emitter in \ls by exploring the consequences of HAWC observational results using analytical tools. This analysis, approximate but suitable for this source due to the large uncertainties involved, is based on the spectral and variability properties of
the UHE emission and is conducted within the framework of IC scattering off  stellar photons. We took $\epsilon=100$~TeV as the reference photon energy to study the orbital variability, a compromise between assuming that the modulation is only real up to $\epsilon\sim 40$~TeV, or that it actually goes beyond $\sim 100$~TeV as INFC photons reach $\approx 208$~TeV while SUPC photons reach just $\approx 118$~TeV, and \citealt{haw24} found $2.2\sigma$ evidence of orbital variability above 100~TeV. The present work strengthens the conclusions of \cite{kha08}, gets deeper into the global implications for the emitter and the accelerator, and proposes an acceleration mechanism based on \cite{der03} and \cite{der12}. The paper is organized as follows. The emitting and the acceleration regions are discussed in Sects.~\ref{em} and \ref{sacc}, respectively, and the suggested model is proposed in Sect.~\ref{mod}. We conclude with Sect.~\ref{conc}.

\section{The UHE photon emitter}\label{em}

While the \hess observational results presented in \cite{aha06} already put important limits on the properties of the VHE emitter and the associated accelerator in \ls \citep[e.g.,][]{kha08}, the recent HAWC
results have yielded tighter constraints, as can be seen in the discussion carried
out by \cite{haw24} in the context of the leptonic scenario. This is particularly true if the emission indeed reaches higher energies around the INFC than around the SUPC. 
To get deeper into the consequences of the HAWC results, we start by
studying the UHE emitting region adopting the framework of stellar IC for the
reasons given above. For those aspects concerning IC scattering, we refer to
\cite{kha14}. As mentioned in Sect.~\ref{intro}, we took $\epsilon=100$~TeV as the photon energy reference. To produce 100~TeV photons, IC scattering off stellar photons with $\epsilon_*\approx 10$~eV takes place deep in the Klein-Nishina regime (KN; i.e., $E\gg m^2_ec^4/\epsilon_\star$), and requires electrons with $E\approx 120$~TeV. 

Anisotropic IC scattering and $\gamma\gamma$ absorption represent two geometrical factors that may influence the orbital phase dependence of the gamma-ray emission, but their influence is strongly suppressed in the UHE regime. Thus, the orbital UHE variability is caused by one or several of these other factors changing along the orbit: hydrodynamics (i.e., escape, adiabatic losses, or heating); intrinsic emission beaming (i.e., Doppler boosting or an anisotropic electron angular distribution); and IC target density. All these effects (including the minor geometric ones) can only be present if the emitter size ($R_{\rm em}$) is not significantly larger than $a$, as otherwise orbital variability would be washed out. Moreover, the mean free path, $\lambda,$ of the highest energy particles in the emitter (at least equal to their gyroradius $r_{\rm g}=E/q{\cal B}_{\rm em}$, where $q$ is the absolute value of the particle charge and \({\cal B}_{\mathrm{em}}\) is the magnetic field strength in the emitter) should fulfill 
$(\lambda/r_{\rm g})r_{\rm g}\lesssim R_{\rm em}\lesssim a$.
Thus, electron confinement requires an emitter magnetic field  
\begin{equation}
{\cal B}_{\rm em}\gtrsim 0.2\,\qty(\frac{\lambda}{r_{\rm g}})\qty(\frac{E}{\rm 120TeV})\qty(\frac{R_{\rm em}}{a})^{-1}\,{\rm G}\,,
\label{bem}
\end{equation}
for $a=2.2\times10^{12}$~cm \citep{cas05}. 

Further constraints can be derived from the spectral slope in the UHE regime. Since for $120$~TeV electrons the KN parameter reaches $E\epsilon_\star/m^2_ec^4\sim 5\times 10^3$, one can adopt the asymptotic relation between the photon and the electron spectra. In particular, the gamma-ray differential distribution $dN_\epsilon/d\epsilon\propto \epsilon^{-2.83\pm0.09^{+0.02}_{-0.05}}$ around INFC \citep{haw24} requires an emitting electron spectrum similar to $dN/dE\propto E^{-2}$ within uncertainties. Synchrotron losses modify the electron injection spectrum ($dQ/dE=dN_{\rm inj}/dEdt$) as follows: 
\begin{equation}
dQ/dE\propto E^{-r}\rightarrow dN/dE\propto E^{-(r+1)}\,\,\,\text{for $r\gtrsim 1$; and}
\end{equation}
\begin{equation}
dQ/dE\propto E^{-r}\rightarrow dN/dE\propto E^{-2}\,\,\,\text{for $r\lesssim 1$}.
\end{equation}
Thus, considering that $dN/dE\propto E^{-2}$, one reaches the conclusion that either $dQ/dE\propto E^{-1}$ or harder, or $dQ/dE\propto E^{-2}$ and the magnetic field is low enough to avoid dominant synchrotron losses. One can compare the synchrotron timescale and the shortest possible escape (or adiabatic) loss time, which are
\begin{equation}
t_{\mathrm{sync}}\approx 3.3\,\qty(\frac{E}{ 120\,\mathrm{TeV}})^{-1}\qty(\frac{{\cal B}_{\mathrm{em}}}{1\,{\rm G}})^{-2}\,\mathrm{s}\,\,\,{\rm and}
\end{equation}
\begin{equation}
t_{\mathrm{esc}}\sim R_{\rm em}/c\approx 73\,\qty(\frac{R_{\mathrm{em}}}{a})\,\mathrm{s}\,, 
\end{equation}
respectively (likely, $v_{\rm em}\sim c$ as particles are marginally confined in the region). Assuming dominant escape (or adiabatic) losses, one obtains the constraint
\begin{equation}
{\cal B}_{\mathrm{em}}\lesssim 0.2\,\qty(\frac{E}{120\,\rm TeV})^{-1/2}\,\qty(\frac{R_{\mathrm{em}}}{a})^{-1/2}\,{\rm G}\,;
\end{equation}
at $E=120$~TeV, only for $R_{\rm em}\sim a$  is this marginally compatible with Eq.~(\ref{bem}).  If orbital variability were limited to $\epsilon\lesssim 40$~TeV, meaning a reference electron energy of $E\approx 50$~TeV, negligible synchrotron losses and the condition $\lambda\lesssim a$ would imply that $\lambda\lesssim 3\,r_{\rm g}$, which is still a very strong confinement condition.
Therefore, assuming negligible synchrotron losses in the emitter is narrowly consistent with $r_{\rm g}\lesssim R_{\rm em}\lesssim a$ and seems to be disfavored (implying a very hard $dQ/dE$; see Sect.~\ref{mod}, however).

To provide further context, one can derive an (conservative) upper limit on the magnetic field strength in the emitter: 
\begin{equation}
{\cal B}_{\mathrm{em}}\lesssim \sqrt{\frac{L_{\rm w}}{v_{\rm em}R_{\rm em}^2}}\approx  3\,\qty(\frac{L_{\mathrm{NT}}}{10^{36}\,{\mathrm{erg/s}}})^{1/2}\qty(\frac{v_{\mathrm{em}}}{c})^{-1/2}\qty(\frac{R_{\mathrm{em}}}{a})^{-1}{\mathrm{G}},
\label{eqb} 
\end{equation}
where $v_{\mathrm{em}}$ is the emitting flow velocity and a spherical geometry is assumed. The  equation was derived taking into account that the Poynting flux cannot overcome the total power of the source ($L_{\rm w}$), and that this power minimum value is that required to feed the NT emission in LS~5039 (i.e., $L_{\mathrm{NT}}\gtrsim 10^{36}$~erg~s$^{-1}$ from the overall MeV-peaked SED; \citealt{col14}). If electrons escaped to the stellar wind region and radiated there, they would find a magnetic field:
\begin{equation}
{\cal B}_{\mathrm{em}}\lesssim \sqrt{\frac{L_{\rm w,*}}{v_{\rm *}R_{\rm em}^2}}\approx 
11\,\qty(\frac{\dot{M}_{\mathrm{*}}}{10^{-7}{\mathrm{M}}_\odot{/\mathrm{yr}}})^{1/2}\qty(\frac{v_{\mathrm{*}}}{2000\,{\mathrm{km/s}}})^{-1/2}\qty(\frac{R_{\mathrm{em}}}{a})^{-1}{\mathrm{G}},
\end{equation}
where $L_{\rm w,*}$ is the stellar wind power, and the wind mass-loss rate ($\dot{M}_{\rm *}$) and velocity ($v_{\rm *}$) were normalized to values typical for \ls \citep{cas05}. These estimates show that, at the energies considered ($E\gtrsim 50$~TeV), \mbox{${\cal B}_{\rm em}$ values} well below equipartition can already lead to dominant synchrotron cooling.

It is informative to derive the synchrotron luminosity of the electrons behind the UHE emission ($L_{\rm sync,UHE}$) under the assumption that $R_{\rm em}\sim a$. We focused on photon energies $\epsilon\gtrsim 40$~TeV ($E\gtrsim 50$~TeV), for which the INFC luminosity detected by HAWC is $L_{\rm HAWC}\approx 2.1\times 10^{32}$~erg~s$^{-1}$ at a 2~kpc distance. The synchrotron luminosity is determined by the ratio of the relevant cooling times:
\begin{equation}
L_{\mathrm{sync,UHE}}\sim \frac{t_{\mathrm{KN}}}{t_{\mathrm{sync}}}L_{\mathrm{HAWC}}\sim 3\times 10^{33}\,\qty(\frac{{\cal B}_{\rm em}}{0.4\,\rm G})^2\,{\mathrm{erg~s}}^{-1}\,,
\end{equation}
where $t_{\rm KN}$ is the KN IC timescale \citep{kha14}: 
\begin{equation}
t_{\rm KN}\sim 7\times 10^2\,\qty(\frac{E}{50\,{\rm TeV}})^{0.7}\,\qty(\frac{\max[R_{\mathrm{em}},a]}{a})^2\,{\mathrm{s}}\,,
\end{equation}
for the star luminosity and temperature \citep{cas05}, and imposing a minimum distance to the star of $a$ (around the location of the compact object). Adopting  ${\cal B}_{\rm em}\sim 3$~G, for instance, $L_{\rm sync,UHE}$ is $\sim 2\times 10^{35}\,{\mathrm{erg~s}}^{-1}$ and the energy of the synchrotron photons from electrons with $E\gtrsim 50$~TeV is $\epsilon\gtrsim 500$~MeV\footnote{This energy is well above the synchrotron burnoff limit, which suggests an accelerator separated from the emitter (see also Sect.~\ref{mod}).}. As shown above, $dN/dE\propto E^{-2}$, which implies a synchrotron SED $\propto \epsilon^{1/2}$ at $\epsilon\gtrsim 500$~MeV. This SED is harder than, and the derived $L_{\rm sync,UHE}$-value similar to, what was observed at those energies \citep[e.g.,][]{had12,col14}, which point to an electron component dominating at lower energies that is different from that producing the UHE emission. This also means that, since under dominant synchrotron losses $L_{\mathrm{NT,UHE}}\sim L_{\mathrm{sync,UHE}}$, in that case $L_{\rm NT,UHE}$
must be lower than $\sim 2\times 10^{35}\,{\mathrm{erg~s}}^{-1}$ (i.e. ${\cal B}_{\rm em}<3$~G) not to violate the observed GeV fluxes.

For the case with negligible synchrotron losses (e.g., ${\cal B}_{\rm em}<0.4$~G for $E=50\,{\rm TeV}$), escape (adiabatic) losses most likely dominate over any radiative cooling (including IC), and thus one obtains (for $R_{\rm em}\sim a$)
\begin{equation}
L_{\mathrm{NT,UHE}}\sim \frac{t_{\mathrm{KN}}}{t_{\mathrm{esc}}}L_{\mathrm{HAWC}}\sim 3\times  10^{33}\,{\mathrm{erg~s}}^{-1}\,,
\end{equation}
which implies that $L_{\mathrm{NT,UHE}}\ll L_{\rm NT}$ and again a different electron component should be behind the brighter emission at lower energies. Increasing $L_{\mathrm{NT,UHE}}$ would require a $R_{\rm em}\ll a$, which is not possible if ${\cal B}_{\rm em}$ is to be low while simultaneously $\lambda\lesssim R_{\rm em}$ (see, however, Sect.~\ref{mod}).

\section{The accelerator}\label{sacc}

Since accelerated electrons may escape the accelerator into a different region where they would produce most of the UHE emission, the properties of the accelerator must be derived separately. Strong constraints on the accelerator magnetic field (${\cal B}_{\rm acc}$) can also be obtained by considering $\lambda\lesssim R_{\rm acc}\lesssim a$, with $R_{\rm acc}$ being the accelerator size. The acceleration timescale for electrons of energy $E$, $t_{\rm acc}$, was defined as
\begin{equation}
t_{\rm acc}\sim \eta_{\rm acc}\,\frac{r_{\rm g}(E)}{c}=\eta_{\rm acc}\frac{E}{q{\cal B}_{\rm acc}c}\,,\,\,\,{\rm with}\,\,\,\eta_{\rm acc}>1\,.
\end{equation}
Here, ${\cal B}_{\rm acc}$ refers to the component of the electromagnetic field doing the acceleration work, which is in fact the electric field depending on the frame of reference or the acceleration mechanism considered.

The constraint $\lambda\lesssim R_{\rm acc}\lesssim a$ implies that, similarly to ${\cal B}_{\rm em}$,
\begin{equation}
{\cal B}_{\rm acc}\gtrsim 0.2\,\qty(\frac{\lambda}{r_{\rm g}})\qty(\frac{E}{120\,\mathrm{TeV}})\qty(\frac{R_{\rm acc}}{a})^{-1}\,{\rm G}\,. 
\end{equation}
Moreover, the condition $t_{\rm acc}\lesssim t_{\rm sync}$ must be fulfilled for electrons to be able to reach an energy $E$, so:
\begin{equation}
t_{\rm acc}\lesssim t_{\rm sync}\rightarrow {\cal B}_{\rm acc}\lesssim 0.2\,\qty(\frac{\eta_{\rm acc}}{1})^{-1}\,\qty(\frac{E}{120\,\rm TeV})^{-2}\,{\rm G}\,;
\end{equation}
for $R_{\rm acc}\sim a$, this is marginally compatible only with $\lambda\sim r_{\rm g}$ and $\eta_{\rm acc}\sim 1$ at $E=120$~TeV, while at $E=50$~TeV, $\lambda\lesssim 3\,r_{\rm g}$ and $\eta_{\rm acc}\lesssim 3$ are required, which are still quite extreme restrictions on particle confinement and the acceleration rate, respectively. All this was already discussed in \cite{kha08} in the context of \hess results (see, e.g., Fig.~2 in that paper), but HAWC results strengthen the observational constraints. We remark that few known astrophysical sources present such demanding requirements for both $t_{\rm acc}$ and $\lambda$. Nonetheless, if synchrotron losses and particle confinement could be decoupled, the requirements would relax significantly. In the next section, we consider an acceleration model based on \cite{der03} and \cite{der12} in which this may happen.

\section{Acceleration model}\label{mod}

Synchrotron cooling and confinement decouple when particles that are roughly isotropic in the laboratory frame (LF) are immersed in a magnetized relativistic outflow. Under these conditions, the flow frame (FF) magnetic field is able to deflect the particles enough for them to acquire a large amount of energy in the LF (as demonstrated in Sects.~\ref{maxen} and \ref{sec:LF}), while the rate of synchrotron cooling in the FF is significantly reduced (as shown in Sect.~\ref{syncloss}). The presence of a bright stellar companion provides a way to bring relativistic particles into the relativistic outflow: LS~5039 is optically thick for gamma rays with energies between $\sim 10$~GeV and 10~TeV interacting with stellar photons on binary scales, so secondary $e^\pm$ pairs (secondaries hereafter) are expected to be injected within the binary \citep[e.g.,][]{dub06b,aha06b}, and thus within the outflow produced by the compact object \citep{der12}. These secondaries, which constitute the seed particles for the acceleration process, are mostly created with energies in the range $\sim 10-100$~GeV, depending on their injection location in the system with respect to the gamma-ray emitter \citep{bos08a}. Secondaries could also be created in the local NT emitter photon fields\footnote{Properly accounting for these local photon fields requires a detailed model of the overall NT emitter, which is beyond the scope of this work.}, 
although we focus on those from $\gamma\gamma$ absorption in the stellar photon field because they reach the highest energies once accelerated. 

\begin{figure}
\includegraphics[width=0.5\textwidth]{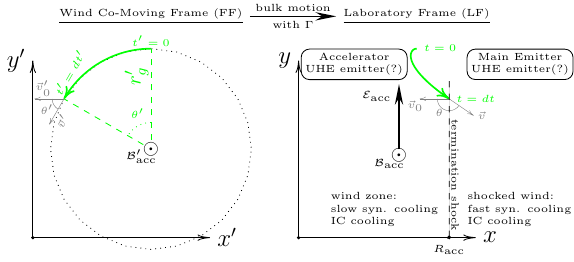}
\caption{Sketch of the acceleration process. An electron gyrates in the $XY$ plane in the FF (left) following a circular trajectory (green line), whereas in the LF (right) it gets quickly accelerated initially by the electric field of the wind, and deflected towards the wind direction of motion. The wind moves in the $X$ direction, and the FF and LF magnetic field and the LF electric field are perpendicular to the flow motion.}
\label{sk}
\end{figure}

To derive a simple analytical estimate of the maximum achievable energies in the described scenario, we assumed a planar wind moving with bulk speed \(v_{\mathrm{w}}\) carrying a perpendicular magnetic field that terminates at a distance $R_{\rm acc}$ from its origin. We assumed that the wind is ultrarelativistic, $\Gamma=1/\sqrt{1-\beta_{\mathrm{w}}^2}\gg1$, where \(\beta_{\mathrm{w}}=v_{\mathrm{w}}/c\). The wind termination would occur in a shock caused by the presence of the stellar wind. Most of the injected secondaries, which are assumed to have an energy $E_0$, initially move in the FF within a cone of half opening angle $\sim 1/\Gamma$ against the wind. Once injected, the secondaries gyrate around the magnetic field, which is ${\cal B}_{\rm acc}$ and ${\cal B}'_{\rm acc}={\cal B}_{\rm acc}/\Gamma$ in the LF and the FF, respectively (primed quantities are in the FF). The secondaries injected in the direction opposite to the flow motion have the highest energy in the FF: $E'=(1+\beta_{\mathrm{w}})\Gamma E_0\approx 2\Gamma E_0$, and a gyroradius
\begin{equation}\label{eq:gyro}
r'_{\rm g}=\frac{E'}{q{\cal B}'_{\rm acc}}=\frac{(1+\beta_{\mathrm{w}})\Gamma^2E_0}{q{\cal B}_{\rm acc}}\approx 2\Gamma^2r_{\rm g,0}\,,
\end{equation}
where both $E'$ and $r'_{\rm g}$ are constant in the FF in a planar wind.
Most of the secondaries injected with energy $E_0$, if moving isotropically in the LF, have similar though slightly lower $E'$ values. For simplicity, we assumed no electric field in the wind FF, that is, the plasma has infinite conductivity. A sketch of the process from the point of view of both the FF and the LF is shown in Fig.~\ref{sk}.

Equation~\eqref{eq:gyro} provides an illustrative way to find an upper limit for the maximum change in the particle energy in the LF. If a particle gyrates for exactly half an orbit starting from an initial orientation against the wind, then its maximum vertical displacement is \(2r'_{\mathrm{g}}\) in the FF; in the LF, the maximum particle vertical displacement is \(2r'_{\mathrm{g}}\) as well. Since in the LF the electric field, $\vec {\cal E}_{\mathrm{acc}}$ , is also perpendicular to the wind bulk speed, the maximum possible energy gain is thus
  \begin{equation}
    \Delta E_{\mathrm{max}} = 2r'_{\mathrm{g}}q{\cal E}_{\mathrm{acc}}=2(1+\beta_{\mathrm{w}})\Gamma^2E_0\beta_{\mathrm{w}}\approx {4\Gamma^2E_0\beta_{\rm w}}\,,
    \label{emaxgyr}
  \end{equation}
  where we used the relation between the electric and the magnetic field strengths \({\cal E}_{\mathrm{acc}}=\beta_{\mathrm{w}} {\cal B}_{\mathrm{acc}}\).

\subsection{Maximum energy}\label{maxen}

With no acceleration size constraints, a particle injected in the direction against the wind bulk velocity keeps gyrating, and its energy in the LF follows a periodic pattern between $E_0$ and $E_0+\Delta E_{\mathrm{max}}$. The larger the $E_0$, the larger the potential drop the particle probes, but the longer the time it requires to gyrate. Therefore, particles can only reach their possible maximum energy if their trajectories are confined within the unshocked wind while they complete half an orbit turn. To determine the maximum energy for particles that cannot gyrate half an orbit before reaching the wind termination shock, one needs to solve the particle motion equation in the wind zone accounting for its size and geometry. This equation can be solved either in the LF or in the FF, and each of these systems has its own advantages: the accelerator size constraints are more clearly defined in the LF; however, solving the particle motion under a uniform magnetic field is trivial in the FF. Therefore, in this work we analyzed the particle gyration angle in the FF and then applied a Lorentz transformation to the LF while taking into account the accelerator size constraint in that frame. This is equivalent to solving the equations of motion directly in the LF. For the sake of completeness, a brief discussion of the effect of the Lorentz force in the LF is also provided in Sect.~\ref{sec:LF}.

The energy acquired in the LF by the secondaries, which are ultrarelativistic, can be derived from
\begin{equation}\label{dsh}
E=\Gamma(1-\beta_{\mathrm{w}}\cos\theta')E'\approx 2\Gamma^2(1-\beta_{\mathrm{w}}\cos\theta')E_0\,,
\end{equation}
where \(\theta'\) is the angle by which the particle is deflected in the FF. At the moment of the particle injection, \(\theta'=0\) and the particle moves in the direction opposite to the wind bulk speed. As shown in Fig.~\ref{sk}, in the FF, the deflection angle, \(\theta'\), is equal to the polar angle swept by the particle. The angle \(\theta\) corresponds to the angle $\theta'$ in the LF (see in Fig.~\ref{sk}), and both angles are related through a Lorentz transformation:
  \begin{equation}\label{eq:angl_LT}
    \tan\theta = \frac{v'\sin\theta'}{\Gamma\qty(v'\cos\theta'-v_{\mathrm{w}})}\,,
  \end{equation}
  where \(v\) and \(v'\) are particle speeds in the LF and FF, respectively, and we accounted for the way these angles are defined (\(v'_x= -v'\cos\theta'\), \(v'_y= -v'\sin\theta'\), \(v_x= -v\cos\theta\), and \(v_y= -v\sin\theta\)).

The increase of the particle energy in Eq.~\eqref{dsh} is consistent with Eq.~(\ref{emaxgyr}) as the particle displacement in the direction of the LF electric field is $r'_{\mathrm{g}}(1-\cos\theta')$.
  The corresponding change of energy is thus 
  \begin{equation}\label{eq:energy_change}
   \Delta E = r'_{\mathrm{g}}(1-\cos\theta')q{\cal E}_{\mathrm{acc}} = (1+\beta_{\mathrm{w}})\Gamma^2E_0\beta_{\mathrm{w}}(1-\cos\theta')\,,
 \end{equation}
which is the same as Eq.~(\ref{emaxgyr}) when $\theta'=\pi$.

The maximum angle $\theta'$ for which the particle is still confined in the wind zone can be obtained from the relation between the distance along the $X$ axis in the LF, $x$, and in the FF, $x'$: 
\begin{equation}
x=\Gamma\qty(x'(t')+\beta_{\rm w} ct')\,.
\label{ttpr}
\end{equation}
In the present scenario and setting the origin of coordinates at the place and moment of the injection of the secondary particle, one has
\begin{equation}
x'=r'_{\rm g}\cos(\theta'+\pi/2)=-r'_{\rm g}\sin\theta'\,,
\label{xpr}
\end{equation}
and $\theta'=v't'/r'_{\rm g}$. In the FF, the particle moves along a circular trajectory while it remains in the wind zone. The three-dimensional shape of this region might be very complicated \citep[in particular, pulsar winds can be strongly anisotropic; see, e.g.,][]{bog02}, so for simplicity we introduced the accelerator size constraint through a parameter with length dimension, $R_{\mathrm{acc}}$, linked to the radius of the wind termination shock. The particle is considered to be within the wind zone for $-R_{\rm acc}\le x\le R_{\rm acc}$.

For relatively small $E_0$ values (e.g., $\lesssim 10$~MeV for $\Gamma\sim 10^3$), secondaries gyrate up to $\theta'\gtrsim \pi/2$, and Eq.~(\ref{dsh}) yields $E\sim\Gamma^2 E_0$ (see Eq.~\ref{emaxgyr} above). For the higher $E_0$ values given above though ($\sim 10-100$~GeV), $\theta'$ can be $\ll 1$. In that case, we adopted a third order Taylor expansion: 
\begin{equation}
\sin\theta'\approx \theta'-\theta'^3/6\,,\,\,\,\cos\theta'\approx 1-\theta'^2/2\,,
\label{3rd}
\end{equation}
which in fact yields results correct within $\sim 10$\% even for $\theta'\sim 1$.

From Eqs.~(\ref{ttpr}), (\ref{xpr}) and (\ref{3rd}), the first order Taylor expansion of $\beta_{\mathrm{w}}\approx 1-1/2\Gamma^2$, the $t'$-$\theta'$ relation, and the fact that in the present case $1/(1-(v'/c)^2)^{1/2}\gg\Gamma$, we derived for $x(\theta')$ 
\begin{equation}
x=\Gamma r'_{\rm g}\theta'\qty(-\frac{1}{2\Gamma^2}+\frac{\theta'^2}{6})\,.
\label{xthpr}
\end{equation}
Since the acceleration cycle ends when the particle reaches \(x=R_{\mathrm{acc}}>0\), which implies \(\theta'>\sqrt{3}/\Gamma\), we can approximate the gyration angle by neglecting the ${\cal O}\qty(1/\Gamma^2)$ term as
\begin{equation}
\theta'\approx \Gamma^{-1}\qty(\frac{3E_{\rm H}}{E_0})^{1/3},
\label{thpr}
\end{equation}
where $E_{\rm H}=q{\cal B}_{\rm acc}x=q{\cal B}_{\rm acc}R_{\rm acc}$ (Hillas limit). This equation together with Eqs.~(\ref{dsh}) and (\ref{3rd}) yields the maximum attainable energy:
\begin{equation}
E_{\rm max}\approx 2.1\,E_{\rm H}\,\qty(\frac{E_0}{E_{\rm H}})^{1/3}\,. 
\label{emmmax}
\end{equation}
Assuming a radial wind and a toroidal ${\cal B}_{\rm acc}$, and taking the wind Poynting flux to be a fraction $\eta_{\rm B}$ of the ultrarelativistic wind total power $L_{\rm w}$, we obtained ${\cal B}_{\rm acc}R_{\rm acc}=(\eta_{\rm B}L_{\rm w}/c)^{1/2}$ and 
\begin{equation}
E_{\rm max}\approx 200\,\qty(\frac{\eta_{\rm B}L_{\rm w}}{3\times10^{36}{\rm erg/s}})^{1/3}\qty(\frac{E_0}{\rm 100GeV})^{1/3}\,{\rm TeV}\,,
\label{emax}
\end{equation}
similar to those required to explain HAWC observations. Higher energies could be achieved for larger $E_0$ values, but the number of secondaries strongly diminishes well above $100$~GeV. We note that, for a final $\theta'<\sqrt{3}/\Gamma$, $x$ in Eq.~(\ref{xthpr}) is negative and $E$ barely grows above $E_0$. This applies in the unlikely event that $E_0$ approaches $E_{\rm H}$, so particles escape almost undeflected in the FF. 

\subsection{Laboratory frame description}\label{sec:LF}

The efficiency of the acceleration process can also be illustrated from the point of view of the LF. Namely, the Lorentz force equation (we focus here on an electron with charge \(-q\)), 
\begin{equation}
\dot {\vec p} = -q\qty({\vec {\cal E}}_{\mathrm{acc}} +
\frac{\vec v}{c}\times\vec {\cal B}_{\rm acc}),
\label{lfor}
\end{equation}
can be directly linked to the energy gain rate, $\dot{E}$. First, $E$ is related to the relativistic three-momentum vector $\vec p$ through 
\begin{equation}
E^2=m^2c^4+c^2{\vec p}^2.
\label{emom}
\end{equation}
If one derives in time both sides of Eq.~(\ref{emom}), it yields
\begin{equation}
E\dot{E}=c^2{\vec p}\dot{\vec p}\,,
\end{equation}
and taking into account that \({\vec p}c^2 = {\vec v}E\), one can write \citep[see, e.g.,][]{1971ctf..book.....L}
\begin{equation}
\dot{E} = \frac{c^2 \vec p}{E}\dot{\vec p}={\vec v}\dot{\vec p}\,.
\end{equation}
Using this relation and ${\vec {\cal E}}_{\rm acc}=-{\vec\beta}_{\mathrm{w}}\cross {\vec {\cal B}}_{\rm acc}$ in Eq.~\ref{lfor} , one obtains
\begin{equation}
\dot{E}={\vec v}\dot {\vec p}=q\,{\vec v}({\vec\beta}_{\mathrm{w}}\times{\vec {\cal B}}_{\mathrm{acc}}-
\frac{{\vec v}}{c}\times{\vec {\cal B}}_{\mathrm{acc}})\,,
\end{equation}
and since ${\vec v}({\vec\beta}_{\mathrm{w}}\times{\vec {\cal B}}_{\mathrm{acc}})=v\beta_{\mathrm{w}} {\cal B}_{\rm acc}\sin\theta$ and ${\vec v}(({\vec v}/c)\times{\vec {\cal B}}_{\mathrm{acc}})=0$,\begin{equation}\label{lforcemin}
\dot{E}=qv\beta_{\mathrm{w}} {\cal B}_{\rm acc}\sin\theta\approx q{\cal B}_{\rm acc}c\sin\theta\,.
\end{equation}
From this expression it can be seen that the most efficient acceleration occurs when \(\theta\approx\pi/2\), which in the FF corresponds to \(\cos\theta'-\beta_{\mathrm{w}}\approx0\), i.e., when \(\theta'\approx 1/\Gamma\) (although, as already explained, particles gain most of their LF energy when $\theta'>\sqrt{3}/\Gamma$).

Equation~(\ref{lforcemin}) can be computed using the relation $\theta'\approx \qty(3q{\cal B}_{\mathrm{acc}}c/E_0\Gamma^3)^{1/3}t^{1/3}$ obtained from the Lorentz transformation for time:
\begin{equation}
  t = \Gamma\qty(t'+\beta_{\mathrm{w}}\frac{x'}{c})\,.
\end{equation}
Using Eq.~\eqref{eq:angl_LT} in the limits \(1/\Gamma\lesssim\theta'\ll1\) and $v'\approx v_{\rm w}\approx c$ one obtains
  \begin{equation}
    \sin\theta \approx \frac{2}{\Gamma\theta'}\,;\,\,\,{\rm i.e.}\,\,\,\sin\theta\approx 2\qty(\frac{E_0}{3q{\cal B}_{\mathrm{acc}}c})^{1/3}t^{-1/3}.
  \end{equation}
Using this last expression, Eq.~(\ref{lforcemin}) can be integrated in $t$ yielding
\begin{equation}
E\approx 3q{\cal B}_{\mathrm{acc}}c\qty(\frac{E_0}{3q{\cal B}_{\mathrm{acc}}c})^{1/3}t^{2/3}.
\end{equation}
For $t\approx R_{\rm acc}/c$, it turns out that the above estimate for the particle energy $E$ is equal to $E_{\rm max}$ in Eq.~(\ref{emmmax}), as expected from the equivalence of both approaches. 

\subsection{Synchrotron losses}\label{syncloss}

Synchrotron losses are negligible in the discussed scenario as the synchrotron timescale in the FF,
\begin{equation}
t'_{\rm sync}\approx
200\,\qty(\frac{\Gamma}{10^3})\,\qty(\frac{E_0}{100\mathrm{GeV}})^{-1}\qty(\frac{{\cal B}_{\mathrm{acc}}}{100{\rm G}})^{-2}\,\mathrm{s}\,, 
\end{equation}
is much longer than $t'$ for typical parameter values:
\begin{equation}
t'\approx 10\,\qty(\frac{R_{\rm acc}}{10^{11}{\rm cm}})^{1/3}\qty(\frac{E_0}{\rm 100GeV})^{2/3}\qty(\frac{\Gamma}{10^3})^{-1/3}\qty(\frac{{\cal B}_{\rm acc}}{\rm 100G})^{-2/3}\,\mathrm{s}\,;
\end{equation}
the normalization ${\cal B}_{\rm acc}$ value of 100~G corresponds to the fiducial values of the wind parameters adopted in Sect.~\ref{maxen} and $\eta_{\rm B}=1$.
Therefore, ${\cal B}_{\rm acc}$ can be high in the lab frame but the accelerated secondaries mostly radiate via IC while they are confined to the ultrarelativistic wind; $t'_{\rm sync}$ can become $<t'$ close to the compact object, but for typical parameter values this region is tiny and does not contribute significantly. The relation  $E_{\rm max}\propto t^{2/3}$ shows that particles accelerate faster initially. Thus, in a more realistic radial wind, particles would acquire most of the energy in a distance shorter than that in which wind expansion becomes noticeable, that is, the adopted planar wind approximation already gives a reasonably accurate result. 

\subsection{Electron spectrum and emitting region}

In the presented acceleration model, particles get a large energy boost in the relativistic outflow, and a very hard injection particle spectrum can be achieved at the wind termination shock. Assuming monoenergetic particles for the sake of simplicity, since energy evolution is $E\propto t^{2/3}$ in the accelerator, the effective energy spectrum of these particles is \(dN_{\mathrm{acc}}/dE\approx (dE/dt)^{-1}\dot{N}_{\pm}\propto E^{1/2}\), where \(\dot{N}_{\pm}\) is the injection rate of secondaries. An \(E^{1/2}\) spectrum of electrons generates a very hard gamma-ray differential spectrum even in the Klein-Nishina regime; \(dN_\epsilon/d\epsilon\propto \epsilon^{-1/2}\), which is incompatible with the HAWC differential spectrum. Nevertheless, the actual shape of $dN_{\mathrm{acc}}/dE$ strongly depends on the secondary spectrum at injection; if the spectrum softens in the region $E_0\sim 100$~GeV, $dN_{\mathrm{acc}}/dE$ will also soften close to $E_{\rm max}$. Given all this, two possible scenarios arise for the accelerator+UHE emitter: 

i) The accelerator features a hard particle spectrum up to $E_{\rm max}$: While still in the wind zone, radiative losses remain negligible, but once particles are injected downstream of the wind termination shock, synchrotron losses render $dN/dE\propto E^{-2}$ for a $dQ/dE\propto E^{-1}$ or harder. The IC emission from these particles is consistent with the HAWC SED, although their corresponding synchrotron emission would not be consistent with the spectrum at lower energies requiring another emitting particle component (or region), as already discussed at the end of Sect.~\ref{em}. This would be compatible with the fact that a significant part of the X-rays and VHE gamma rays seem to come from a region with a size close to but $\gtrsim a$ \citep[e.g.,][]{bos07,kha08,bos08b,szo11}.

ii) A soft particle spectrum at $\lesssim E_{\rm max}$: The highest energy particles may follow $dN_{\mathrm{acc}}/dE\propto E^{-2}$ in the wind itself and be behind the HAWC emission. The emission below HAWC energies may be produced by the wind accelerated particles with lower energies once these particles reach the wind termination shock, or further downstream, as suggested in previous works. 

In both cases (i and ii), the accelerator should not violate the HAWC fluxes, which means that
\begin{equation}
R_{\mathrm{acc}}\lesssim \qty(\frac{L_{\mathrm{HAWC}}}{L_{\mathrm{NT,UHE}}})\,t_{\mathrm{KN}}\,c\,.
\label{dst}
\end{equation}
For instance, taking $L_{\mathrm{NT,UHE}}\sim 10^{35}$~erg~s$^{-1}$ ($\lesssim 0.1\,L_{\rm NT}$) and $E\sim 100$~TeV, one obtains $R_{\mathrm{acc}}\lesssim 10^{11}$~cm. In fact, even $L_{\rm NT,UHE}\sim L_{\rm NT}$ would be allowed due to the relaxation on the confinement constraint, as now $R_{\rm acc}$ can be significantly smaller than $a$.
The distance derived in Eq.~(\ref{dst}) is to be compared to that for which the relativistic and the stellar wind are in pressure balance, whose smallest value from the compact object is
\begin{equation}
R_{\rm eq}\sim 3\times 10^{11}\,\frac{(\eta/0.03)^{1/2}}{1+(\eta/0.03)^{1/2}}\,{\rm cm}\,, 
\end{equation}
where
\begin{equation}
\eta=\frac{L_{\rm NT}}{\dot{M}_{\mathrm{*}}v_{\mathrm{*}}c}\approx 0.03\,\qty(\frac{L_{\rm NT}}{10^{36}{\rm cm/s}})\qty(\frac{\dot{M}_{\rm *}}{10^{-7}{\rm M}_\odot/{\rm yr}})^{-1}\qty(\frac{v_{\rm *}}{2000{\rm km/s}})^{-1}
\end{equation}
is the minimum ratio of momentum rates between the ultrarelativistic and the stellar wind. Given that observational constraints make it difficult to reduce $\eta$ much further, there seems to be a disparity between $R_{\rm acc}$ and $R_{\rm eq}$, which may be explained by the ultrarelativistic wind being slowed down and heated due to pair-loading \citep{der12} before full termination against the stellar wind (see alternatively Bosch-Ramon \& Barkov, in prep.). This process would strongly affect the particle evolution and emission in the region where it becomes important, but its proper characterization needs detailed calculations.

\subsection{Mechanism comparison}

The acceleration rate in the considered scenario can be roughly estimated from 
\begin{equation}
\dot{E}_{\rm acc}\sim E_{\rm max}/t\approx 0.1\,\qty(\frac{E_0/E_{\rm H}}{10^{-4}})^{1/3}\,q{\cal B}_{\rm acc}c\,.
\label{edot}
\end{equation}
For comparison, in the most optimistic case, relativistic magnetic reconnection can provide $\dot{E}_{\rm acc,rec}\sim q{\cal B}_{\rm acc}c$, which would allow the attainment of $E_{\rm max,rec}\sim E_{\rm H}\sim 10\,E_{\rm max}$ for the normalization parameter values in Eq.~(\ref{edot}); synchrotron losses may also be small in this case if the guiding magnetic field is negligible, as it is the electric field that accelerates the particles. However, for $E_{\rm max,rec}\gtrsim E_{\rm max}$, the current sheet region should be coherent on a scale $\gtrsim 0.1\,R_{\rm acc}$ and the guiding magnetic field should be small; these conditions may be difficult to realize. Regarding diffusive acceleration processes, one must keep in mind that those mechanisms would struggle to reach $E_{\rm max}\sim 100$~TeV for the reasons already discussed in Sects.~\ref{em} and \ref{sacc}. 

\section{Conclusions}\label{conc}

Recent results from HAWC show that \ls features a relatively hard SED and orbital
variability up to UHE. Most of this emission, like the VHE emission, is
produced in the INFC. This, together with the system properties, strongly suggests the same radiation mechanism in both energy bands, most
likely IC scattering off stellar photons. The exceedingly high energies of the
electrons behind the UHE emission, and the emission orbital modulation, imply very
extreme conditions of the accelerator and the emitter. Even under the most
conservative assumption that variability only affects photon energies up to
$\approx 40$~TeV, synchrotron losses must be very small in the accelerator, but
the corresponding magnetic field can hardly confine the particles in the region. A
potential solution to this problem is a version of the converter mechanism
proposed by \cite{der12} in the context of a high-mass gamma-ray binary hosting an ultrarelativistic outflow. To obtain the maximum electron energies, we proposed a simplified model
based on that mechanism that significantly alleviates the tensions, as synchrotron
cooling is reduced without having to decrease the magnetic field. Detailed,
thorough studies are nevertheless needed to properly understand the process
quantitatively; this would better characterize the properties of the injected
secondaries, the accelerated particle spectrum, the connection between the UHE and
the broadband NT emitter, and the specific UHE variability cause(s). 

In the proposed scenario, \ls hosts a compact and magnetized
ultrarelativistic outflow that powers all the NT activity in the system. This
outflow, which likely originates from a pulsar, also accelerates the electrons behind the UHE emission. The UHE emitter, which is unlikely to coincide with the broadband NT emitter, may be the
ultrarelativistic outflow itself or be located beyond its termination shock. The ultrarelativistic outflow region could be
significantly smaller than the one bounded by the stellar wind, which may be
explained for instance by braking and heating of the outflow due to pair-loading. 

Other
processes and regions in the source besides the ultrarelativistic outflow are potential accelerators and emitters that contribute to the overall NT radiation: i) magnetic reconnection in the outflow or at its termination; ii) the slowed down and heated outflow and its termination shock and the regions beyond; iii) the  Coriolis shock; and iv) the large-scale interaction with the interstellar medium (ISM) \citep[for previous work of modeling of some of these regions in the context of \ls, see, e.g.,][]{tak09,bos11,dub15,del15,mol20,hub21,kef24}.
Under very ideal
conditions, option (i) may also be a feasible mechanism to explain HAWC results. On the other hand,
options (ii) and (iii) may produce variable UHE emission, but are severely
affected by the extreme requirements on the particle acceleration rate and confinement
discussed here. Option (iv), the large-scale ISM interaction, may provide
steady multiwavelength NT emission \citep[e.g., extended X-ray emission has been found in
LS~5039;][]{dur11}, and its contribution to the Galactic cosmic rays may not be
negligible. In all the regions where stellar wind mixing had already occurred \citep[e.g.,][]{bos11,bos12}, protons may also be accelerated. Among the regions discussed, region (iv) is favored as a potential cosmic ray contributor, as adiabatic losses should affect the particles propagating from the binary scales to the ISM. This is particularly hard to avoid under conditions of strong particle confinement.

\begin{acknowledgements}
We want to thank the referee for constructive and useful comments that helped to improve the manuscript. 
We acknowledge the important role of Evgeny Derishev through long and fruitful discussions that contributed to shape the ideas behind this work.
VB-R acknowledges financial support from the State Agency for Research of the Spanish Ministry of Science and Innovation under grant PID2022-136828NB-C41/AEI/10.13039/501100011033/ERDF/EU and through the Unit of Excellence Mar\'ia de Maeztu awards to the Institute of Cosmos Sciences (CEX2019-000918-M; CEX2024-001451-M), and from Departament de Recerca i Universitats of Generalitat de Catalunya through grant 2021SGR00679. V.B-R. is Correspondent Researcher of CONICET, Argentina, at the IAR. DK acknowledges support by RSF grant No. 24-12-00457.
\end{acknowledgements}

\bibliographystyle{aa} 
\bibliography{ref} 
\end{document}